\documentclass[11pt]{article}
\usepackage{abstract}
\usepackage[a4paper,margin=1in]{geometry}
\usepackage{amsmath,amssymb}
\usepackage{microtype}
\usepackage{enumitem}
\usepackage{hyperref}
\usepackage{booktabs} 
\usepackage{graphicx}
\hypersetup{
  colorlinks=true, linkcolor=blue, citecolor=blue, urlcolor=blue,
  pdftitle    ={An intersectional informational data valuation theory},
  pdfkeywords ={Data pricing, Mutual information, Externalities, Intersectionality, Privacy economics}
}

\newcommand{\I}{\mathrm{I}}

\begin{document}
\title{\textbf{Intersectional Data and the Social Cost of Digital Extraction: A Pigouvian Surcharge}}
\author{Eduardo C. Garrido-Merchán\\ 
    Universidad Pontificia Comillas \\
    Instituto de Investigación Tecnológica \\
    \texttt{ecgarrido@comillas.edu}}        
\date{January 2026}
\maketitle

\begin{abstract}
Contemporary digital capitalism relies on the large-scale extraction and commodification of personal data. Far from revealing isolated attributes, such data increasingly exposes intersectional social identities formed by combinations of race, gender, disability, and other protected characteristics. This process generates a structural privacy externality: while firms appropriate economic value through profiling, prediction, and personalization, individuals and social groups bear diffuse costs in the form of heightened social risk, discrimination, and vulnerability. This paper develops a formal political–economic framework to internalize these externalities by linking data valuation to information-theoretic measures. We propose a pricing rule based on mutual information that assigns monetary value to the entropy reduction induced by individual data points over joint intersectional identity distributions. Interpreted as a Pigouvian-style surcharge on data extraction, this mechanism functions as an institutional constraint on the asymmetric accumulation of informational power. A key advantage of the approach is its model-agnostic character: the valuation rule operates independently of the statistical structure used to estimate intersectional attributes, whether parametric, nonparametric, or machine-learned, and can be approximated through discretization of joint distributions. We argue that regulators can calibrate this surcharge to reflect contested social values, thereby embedding normative judgments directly into market design. More broadly, the framework reframes privacy not as an individual preference to be traded, but as a collective condition shaped by power relations in digital markets. By formalizing the social cost of intersectional data extraction, the proposed mechanism offers both a corrective to market failure and a redistributive institutional shield for vulnerable groups under conditions of digital asymmetry.
\end{abstract}

\bigskip

\section{Introduction: Digital Extraction, Intersectional Harm, and Social Cost.}
Digital platforms routinely monetise behavioural traces generated in the course of everyday life. While such data is typically traded at negligible or zero market prices, its disclosure often entails substantial social costs that remain unaccounted for. Seemingly innocuous signals, such as opening a fitness application at a fixed early-morning hour, can sharply increase the inferred probability that an individual belongs to a particular intersection of age, gender, and ability status. Through such inferences, personal data contributes not merely to prediction and personalisation, but to the reconstruction of protected social identities. Conventional pricing mechanisms in data markets systematically ignore this externality, resulting in an overproduction of privacy harm.

This mismatch between private gain and social cost is not a technical oversight but a structural feature of contemporary digital capitalism. As argued by critical political economists, large platforms extract value not primarily from wage labour, but from the continuous appropriation and processing of human behavioural data \cite{Zuboff2019}. This form of extraction is asymmetrical: individuals supply data under conditions of weak bargaining power, limited transparency, and constrained exit, while firms retain exclusive control over the inferential infrastructures that transform trivial signals like location pings, typing rhythms, or even silence, into economically valuable predictions. The result is a quiet process of dispossession, facilitated by default permissions, dark patterns, and the absence of meaningful opt-out mechanisms, disproportionately affecting those with limited digital literacy or institutional power \cite{Eubanks2018}.

The distributive consequences of this process are best understood through the lens of intersectionality. Social harms arising from data extraction do not accumulate additively across isolated attributes, but compound across overlapping identities structured by race, gender, class, and disability amongst others \cite{Crenshaw1989}. A growing empirical literature documents that automated inference and decision systems disproportionately disadvantage women of colour, disabled individuals, and other multiply marginalised groups \cite{Noble2018,Barocas2019}. Crucially, these same populations often generate data under conditions of heightened exposure—such as public assistance platforms, gig economy applications, or default mobile infrastructures in the Global South—where consent is weak and surveillance is dense. As a consequence, marginalised subjects both receive fewer benefits from data-driven services and bear a disproportionate share of their algorithmic risks.

From the standpoint of political economy, these dynamics are closely linked to the concentration of informational power in a small number of dominant platforms. Economies of scale and scope in data aggregation allow large firms to infer protected attributes with increasing accuracy and declining marginal cost, reinforcing entry barriers and enabling forms of de facto cartelisation. This concentration further widens the gap between private returns and social harm, violating core welfare-economic principles regarding externalities and market failure \cite{Stiglitz2000}. In such settings, Pigouvian theory offers a classical corrective: when market transactions impose unpriced social costs, those costs must be internalised through corrective pricing mechanisms.

Existing approaches to data valuation, however, fall short of this task. Early work in privacy economics focused on individuals’ willingness to pay for privacy or willingness to accept compensation for disclosure \cite{Acquisti2005}, implicitly treating privacy as a matter of subjective preference. More recent contributions propose data pricing schemes based on marginal predictive value, Shapley approximations, or consumer surplus in stylised data markets \cite{Bergemann2015,Ghorbani2020,Jia2019}. While technically sophisticated, these models typically assume homogeneous agents and abstract from the social position of data subjects. Parallel strands of research in algorithmic fairness and differential privacy aim to mitigate disparate impact through technical constraints \cite{Feldman2015,Dwork2014}, yet rarely translate fairness concerns into economic prices capable of reshaping market incentives. Only a small number of recent works explicitly consider intersectional dimensions of data valuation \cite{Kattan2022}, and none develop a general economic rule linking price to the social cost of intersectional inference.

This paper addresses that gap by proposing a Pigouvian surcharge on digital data extraction grounded in information theory. We take the perspective of a regulator or, equivalently, of a platform operating under regulatory constraints, seeking an implementable and quantitative rule for internalising privacy externalities. The central question is straightforward: how should the price of a single datum reflect its capacity to reveal a protected intersectional profile of the data subject? Our answer rests on a simple but powerful observation. The expected reduction in uncertainty about a protected profile induced by observing a datum is captured by the mutual information between the two. Interpreted economically, this reduction in entropy measures the extent to which a datum contributes to socially costly inference.

By tying data prices directly to mutual information, the proposed surcharge transforms privacy from a qualitative norm into a measurable economic quantity, while remaining agnostic to the specific statistical models used for inference. More importantly, it reframes data valuation as a problem of political economy rather than individual choice: the price of data reflects not subjective preferences, but the social cost of extracting predictive power over vulnerable identities. In doing so, the paper contributes a formal yet explicitly normative intervention aimed at constraining the asymmetric accumulation of informational power in digital markets.

\section{Data, Power and Privacy Beyond Individual Preferences}

Standard economic accounts of privacy in digital markets often begin from individual choice: users are assumed to trade data for services, and privacy is framed as a preference that can be priced through willingness-to-pay or willingness-to-accept compensation \cite{Acquisti2005}. While analytically convenient, this framing obscures the political economy of contemporary data markets. The central problem is not merely that individuals misprice privacy, but that the production and appropriation of informational value is organised through structural asymmetries of power.

First, participation is rarely meaningfully voluntary. The conditions under which data is surrendered are shaped by default permissions, contractual opacity, and manipulative interface design in the shape of dark patterns that transform consent into a procedural fiction rather than a bargain between equals \cite{Eubanks2018}. Second, even when a user can observe that a datum $X$ is collected, they typically cannot observe what $X$ enables: downstream inferential pipelines that reconstruct protected attributes and intersectional profiles from trivial behavioural traces. Inference thus converts everyday life into a source of predictive surplus, while the risks of discrimination and social sorting remain external to the platform's balance sheet.

This is why privacy harm is not well captured by the language of individual preference alone. It is a collective and distributive condition: the accumulation of predictive capacity by dominant firms reconfigures social vulnerability, particularly for those located at multiply marginalised intersections \cite{Crenshaw1989,Noble2018,Barocas2019}. In the terms of critical political economy, platforms do not merely use data; they extract and concentrate informational power, a stock of classification and prediction that can be deployed for profiling, behavioural steering, and the segmentation of populations \cite{Zuboff2019}. The gains of this extraction are privately appropriated, while its harms are dispersed, often disproportionately borne by groups with weak bargaining power and limited institutional protection.

From a regulatory standpoint, this diagnosis shifts the object of policy. If the core harm is an externality produced by inferential extraction under asymmetric power, then notice-and-consent mechanisms are structurally insufficient: they treat privacy as a contract term when it is better understood as a condition shaped by market structure and institutional constraint. Classical welfare economics offers a familiar corrective where externalities are present: Pigouvian pricing aligns private incentives with social cost by attaching a surcharge proportional to marginal harm \cite{Stiglitz2000}. The challenge is to define an implementable measure of harm that captures intersectional inference rather than merely observable disclosure.

The remainder of the paper develops such a measure and derives a corresponding Pigouvian surcharge. Our aim is not to propose a moral appeal to better corporate behaviour, but an institutional mechanism that makes intersectional informational extraction economically legible and thus regulable. By pricing the social cost of inferential power, the surcharge operates as a constraint on accumulation: it internalises harms that digital capital presently externalises, and it opens a space for explicit political choice in calibrating how much society is willing to tolerate the commodification of protected identity.

\section{Intersectional Data and Information-Theoretic Valuation}

Intersectional theory emphasises that social vulnerability arises not from isolated attributes but from their interaction \cite{Crenshaw1989}. In data-driven systems, this insight has a direct analytical implication: privacy risk is not additive across protected categories, but emerges from their joint configuration. Assessments that focus on marginal leakage about gender, race, or disability in isolation systematically understate the exposure faced by individuals located at vulnerable intersections. What matters politically and economically is the extent to which data enables the reconstruction of an entire protected profile.

To formalise this insight, we represent the protected intersectional identity of a data subject as a multivariate random variable $S = (S_1, S_2, \ldots, S_m)$ where each component corresponds to a protected attribute or dimension of social classification. A datum $X$, such as a timestamp, device feature, or behavioural trace, generates social harm insofar as it reduces uncertainty about $S$. From an information-theoretic perspective, this reduction is naturally measured by the mutual information $\I(X;S)$, which captures expected entropy loss over the full joint distribution of protected attributes. Crucially, this measure respects the intersectional structure of vulnerability: it assigns weight not only to marginal disclosures but to higher-order combinations that often underpin discrimination and social sorting.

This choice has two political–economic advantages. First, it avoids privileging any single protected dimension ex ante, treating intersectional exposure as a structural property of inference rather than a list of sensitive fields. Second, it renders privacy harm commensurable across contexts by expressing it in units of uncertainty reduction, independently of the specific statistical model used for prediction. Whether inference is performed through parametric estimation, machine learning, or nonparametric methods, the social cost of extraction is mediated through the same observable quantity: informational gain about protected identity.

Building on this representation, we articulate a small set of normative requirements that any defensible pricing mechanism for intersectional data extraction should satisfy. These requirements are not derived from a particular moral doctrine, but from the logic of Pigouvian correction applied to informational externalities. In brief, a socially fair price attached to a datum must depend exclusively on the extent to which it enables intersectional inference; it must assign no surcharge where no such inference is possible; it must increase monotonically with informational exposure; and it must scale additively across independent disclosures. Together, these conditions formalise a minimal conception of justice in data valuation: one that prices harm where harm is generated, and remains silent where it is not.

Under these requirements, the valuation problem admits a unique solution. The social cost associated with a datum increases linearly with the mutual information it reveals about the protected profile. Formally, the socially relevant price of accessing $X$ can be written as
\[
V(X) = c_p + \lambda\,\I(X;S),
\]
where $c_p$ denotes the internal marginal cost of producing or processing the datum, and $\lambda>0$ is a policy parameter expressing society’s valuation of intersectional privacy. This expression is not merely a technical convenience: it is the informational analogue of a Pigouvian surcharge. It translates the extraction of predictive power into a monetary liability, making the social cost of intersectional inference explicit and, therefore, regulable.

By grounding valuation in mutual information, the surcharge operates independently of individual preferences and contractual consent. What is priced is not the act of disclosure as such, but the contribution of data to the accumulation of informational power over protected identities. In this sense, the pricing rule formalises an institutional response to a structural imbalance: it constrains the private appropriation of predictive surplus by forcing platforms to internalise the social cost of the intersectional knowledge they extract.

\section{A Pigouvian Surcharge on Digital Data Extraction}

The valuation rule derived in the previous section admits a direct political–economic interpretation. The term $\lambda\,\I(X;S)$ functions as a Pigouvian surcharge imposed on the extraction of intersectional information. Just as environmental taxation internalises the social cost of pollution, this surcharge internalises the social harm generated by predictive inference over protected identities. What is being taxed is not data collection per se, but the contribution of a datum to the accumulation of informational power.

From a welfare-economic perspective, the rationale is straightforward. In unregulated data markets, firms face only the private marginal cost of data production, $c_p$, while the social cost associated with profiling, discrimination, and loss of autonomy remains externalised. As a result, informational extraction is pursued beyond the socially optimal level. By attaching a monetary charge proportional to $\I(X;S)$, the surcharge forces firms to confront the full marginal cost of their inferential activities. Data is then exploited only where its private benefit exceeds its combined private and social cost, restoring a basic condition of allocative efficiency.

At the same time, the surcharge has an explicitly distributive dimension. The harms captured by $\I(X;S)$ are not evenly distributed across the population. Intersectional inference disproportionately affects individuals situated at socially vulnerable positions, those whose identities are historically subject to surveillance, classification, and exclusion. In this sense, the surcharge does more than correct a market failure: it operates as a redistributive mechanism that counteracts the asymmetric appropriation of informational surplus. Where platforms currently extract predictive value without compensation, the surcharge converts inference into a monetised liability, opening the possibility of restitution to affected subjects or collective compensation through public funds.

This logic contrasts sharply with consent-based regulatory approaches. Notice-and-consent frameworks treat privacy loss as a matter of individual choice, implicitly assuming symmetric information and bargaining power. The Pigouvian surcharge, by contrast, recognises that the relevant harm arises downstream of consent, through inference that is opaque to the data subject. Pricing informational extraction therefore shifts regulatory attention from formal agreement to structural outcome: what matters is not whether a datum was agreed to, but what it enables in terms of social classification and control.

The policy parameter $\lambda$ plays a central role in this architecture. It expresses society’s valuation of intersectional privacy and can be calibrated to reflect contested normative priorities. A low value of $\lambda$ tolerates extensive profiling in exchange for innovation and efficiency gains; a high value constrains extraction and shifts incentives toward privacy-preserving design. Importantly, this choice is irreducibly political. There is no technocratic value of $\lambda$ that can be deduced from the model itself. The surcharge thus embeds democratic judgment directly into market pricing, making explicit what is often hidden behind claims of neutrality in algorithmic systems.

From the standpoint of critical political economy, the surcharge can be read as a limit imposed on the accumulation of informational capital. Digital platforms derive durable power not merely from ownership of data, but from their capacity to transform data into predictive and classificatory authority over populations. By increasing the cost of extracting intersectional knowledge, the surcharge raises the price of this accumulation process. It does not abolish informational capitalism, but it constrains its most extractive dynamics, particularly where they target vulnerable groups whose identities function as raw material for prediction.

Finally, the surcharge is institutionally flexible. Revenues may be directed toward individual compensation, collective redress mechanisms, or general public funds supporting social protection. What matters is that informational harm becomes visible and priced, rather than silently absorbed by those least able to bear it. In this sense, the Pigouvian surcharge functions not only as a corrective instrument, but as a social shield: it reintroduces political accountability into data markets by transforming opaque inference into an object of economic and regulatory scrutiny.

\section{Discussion: Market Failure, Structural Power, and the Limits of Pricing}

The Pigouvian surcharge proposed in this paper is intentionally modest in its ambition. It does not claim to resolve the full set of power asymmetries that structure digital capitalism, nor to eliminate the commodification of personal data. Rather, it operates within existing market institutions, seeking to correct a specific and measurable failure: the systematic externalisation of social harm arising from intersectional inference. Acknowledging these limits is essential to avoid both technocratic overreach and misplaced critique.

From a standard welfare-economic perspective, the surcharge addresses a negative externality. However, in the context of data capitalism, the externality is inseparable from structural power. Informational extraction is not a marginal deviation from an otherwise competitive equilibrium; it is a central mechanism through which dominant platforms consolidate and reproduce their position. Pricing informational harm therefore cannot be understood as a neutral efficiency correction alone. It intervenes in a process of accumulation that is historically and institutionally specific.

A Marxian reading makes this point explicit. Digital platforms accumulate what may be termed \emph{informational capital}: a stock of predictive and classificatory capacity that enables the surveillance, segmentation, and behavioural steering of populations. This capital is produced through the continuous appropriation of behavioural traces generated in everyday life, often under conditions where participation is compulsory or unavoidable. The surplus extracted from this process is not returned to data subjects, whose identities become inputs into profit-generating inference without remuneration or control. Intersectional identities, in particular, function as high-yield sites of extraction, as their statistical distinctiveness increases predictive value while social vulnerability reduces resistance.

Seen in this light, the Pigouvian surcharge functions as a partial counter-move. By attaching a monetary cost to the extraction of intersectional information, it raises the price of accumulating informational capital and thereby constrains its most exploitative dynamics. Yet it does not abolish the underlying relations of production. Firms may still find it profitable to pay the surcharge, especially where predictive value is high. This is not a flaw of the mechanism, but a reflection of its scope: pricing can discipline accumulation, but it cannot substitute for broader institutional transformation.

This observation points to a second limitation. Because the surcharge operates through price signals, its effectiveness depends on the responsiveness of firms to cost increases and on the regulatory capacity to enforce accurate measurement of informational leakage. In highly concentrated markets, dominant platforms may possess sufficient market power to absorb or pass on the surcharge without meaningful behavioural change. Moreover, informational opacity remains a persistent challenge. Without robust audit rights and transparency obligations, the estimation of $\I(X;S)$ risks becoming a contested or strategically manipulated exercise.

These limitations, however, should not be mistaken for reasons to abandon pricing altogether. On the contrary, they highlight the importance of situating Pigouvian instruments within a broader political–economic strategy. Pricing makes harm visible; it converts diffuse and often invisible social costs into explicit economic quantities that can be debated, contested, and adjusted. In doing so, it creates institutional footholds for further intervention. A surcharge that proves insufficient can be increased; a pricing scheme that reveals persistent harm can justify more stringent restrictions or outright prohibitions.

Importantly, the framework also resists a common critique of data regulation: that formal mechanisms necessarily depoliticise harm by translating it into technical metrics. In this case, the opposite holds. The parameter $\lambda$ embodies an explicit political choice about how much intersectional harm society is willing to tolerate. Far from eliminating normative conflict, the pricing rule foregrounds it, embedding distributive judgment directly into market design. This stands in contrast to consent-based regimes, which often mask power relations behind procedural formalities.

In sum, the Pigouvian surcharge should be understood neither as a technocratic fix nor as a comprehensive solution. It is a deliberately partial intervention that targets a concrete mechanism of exploitation within digital capitalism: the unpriced extraction of intersectional information. Its value lies not in its completeness, but in its capacity to disrupt the invisibility of informational harm and to impose economic friction on practices that currently operate without constraint. As such, it complements, rather than replaces, broader struggles over data governance, labour, and social justice in the digital economy.

\section{Conclusion}

This paper has argued that contemporary data markets systematically underprice the social harm generated by predictive inference over protected, intersectional identities. While digital platforms monetise behavioural traces at negligible marginal cost, the inferential power derived from those traces produces diffuse yet consequential harms: profiling, discrimination, and the reinforcement of structural vulnerability. These harms are not adequately captured by prevailing frameworks that treat privacy as an individual preference or a contractual choice.

By integrating intersectionality theory with information-theoretic measurement, we have proposed a principled way to make such harms economically legible. Representing protected identity as a joint distribution and measuring disclosure through mutual information allows privacy risk to be understood as a collective, non-additive phenomenon. On this basis, we derived a simple Pigouvian surcharge that links the price of a datum to the expected reduction in uncertainty about an individual’s intersectional profile. The resulting pricing rule is model-agnostic, operationally tractable, and explicitly normative: it embeds political judgment directly into market design through the calibration of a single policy parameter.

Crucially, the proposed surcharge is not framed as a technocratic solution to a moral problem. It is an institutional intervention aimed at correcting a specific mechanism of exploitation within digital capitalism: the unpriced extraction of informational power from socially embedded identities. By forcing firms to internalise the social cost of intersectional inference, the surcharge introduces economic friction into practices that currently operate without constraint, particularly where they target vulnerable populations.

At the same time, we have emphasised the limits of pricing as a regulatory strategy. Informational extraction is deeply entangled with market concentration, opacity, and structural power. A Pigouvian surcharge cannot, on its own, dismantle these relations. Its significance lies instead in rendering harm visible, contestable, and subject to democratic calibration. In this sense, pricing operates not as an endpoint, but as a lever: it opens institutional space for stronger forms of governance, redistribution, or prohibition where necessary.

More broadly, this framework reframes privacy from a matter of individual consent to a question of political economy. What is at stake is not merely who agrees to share data, but who bears the cost of prediction and who accumulates its benefits. By treating intersectional identity as a site of social value extraction rather than a neutral input, the proposed approach aligns economic regulation with long-standing critiques of exploitation and dispossession in capitalist systems.

Future work may extend this model by differentiating surcharges across downstream uses of data such as political profiling, targeted advertising, or automated decision-making or by embedding information-based pricing within broader regimes of data ownership and collective rights. Whatever the institutional form, the central insight remains: as long as predictive power over protected identities can be accumulated without cost, data markets will continue to reproduce inequality. Making that power costly is a necessary step toward reclaiming political control over the informational foundations of social life.

\bibliography{main}

@book{Zuboff2019,
  author    = {Shoshana Zuboff},
  title     = {The Age of Surveillance Capitalism: The Fight for a Human Future at the New Frontier of Power},
  year      = {2019},
  publisher = {PublicAffairs}
}

@book{Eubanks2018,
  author    = {Virginia Eubanks},
  title     = {Automating Inequality: How High-Tech Tools Profile, Police, and Punish the Poor},
  year      = {2018},
  publisher = {St. Martin's Press}
}

@book{Noble2018,
  author    = {Safiya Umoja Noble},
  title     = {Algorithms of Oppression: How Search Engines Reinforce Racism},
  year      = {2018},
  publisher = {NYU Press}
}

@book{Barocas2019,
  author    = {Solon Barocas and Moritz Hardt and Arvind Narayanan},
  title     = {Fairness and Machine Learning},
  year      = {2019},
  note      = {Available at \url{http://fairmlbook.org}}
}

@article{Stiglitz2000,
  author  = {Joseph E. Stiglitz},
  title   = {The Contributions of the Economics of Information to Twentieth Century Economics},
  journal = {The Quarterly Journal of Economics},
  volume  = {115},
  number  = {4},
  pages   = {1441--1478},
  year    = {2000}
}

@article{Crenshaw1989,
  author  = {Kimberlé Crenshaw},
  title   = {Demarginalizing the Intersection of Race and Sex: A Black Feminist Critique of Antidiscrimination Doctrine, Feminist Theory and Antiracist Politics},
  journal = {University of Chicago Legal Forum},
  year    = {1989}
}

@article{Acquisti2005,
  author  = {Alessandro Acquisti and Jens Grossklags},
  title   = {Privacy and Rationality in Individual Decision Making},
  journal = {IEEE Security \& Privacy},
  volume  = {3},
  number  = {1},
  pages   = {26--33},
  year    = {2005}
}

@inproceedings{Ghorbani2020,
  author    = {Amirata Ghorbani and James Zou},
  title     = {Data Shapley: Equitable Valuation of Data for Machine Learning},
  booktitle = {Proceedings of the 37th International Conference on Machine Learning (ICML)},
  year      = {2020}
}

@inproceedings{Jia2019,
  author    = {Ruiqi Jia and Dorsa Sadigh and Moritz Hardt and Sham M. Kakade and Tengyu Ma},
  title     = {Towards Efficient Data Valuation Based on the Shapley Value},
  booktitle = {Proceedings of the 22nd International Conference on Artificial Intelligence and Statistics (AISTATS)},
  year      = {2019}
}

@article{Bergemann2015,
  author  = {Dirk Bergemann and Alessandro Bonatti and Alex Smolin},
  title   = {The Economics of Social Data},
  journal = {RAND Journal of Economics},
  volume  = {46},
  number  = {3},
  pages   = {501--524},
  year    = {2015}
}

@inproceedings{Feldman2015,
  author    = {Michael Feldman and Sorelle A. Friedler and John Moeller and Carlos Scheidegger and Suresh Venkatasubramanian},
  title     = {Certifying and Removing Disparate Impact},
  booktitle = {Proceedings of the 21th ACM SIGKDD International Conference on Knowledge Discovery and Data Mining (KDD)},
  year      = {2015}
}

@article{Dwork2014,
  author  = {Cynthia Dwork and Aaron Roth},
  title   = {The Algorithmic Foundations of Differential Privacy},
  journal = {Foundations and Trends in Theoretical Computer Science},
  volume  = {9},
  number  = {3--4},
  pages   = {211--407},
  year    = {2014}
}

@inproceedings{Kattan2022,
  author    = {Ahmed Kattan and Manasi Vartak and Hoda Heidari},
  title     = {Achieving Fairness via Data Valuation},
  booktitle = {Proceedings of the ACM Conference on Fairness, Accountability, and Transparency (FAccT)},
  year      = {2022}
}
\bibliographystyle{acm}

\end{document}